\documentclass{aastex}
\usepackage{emulateapj5}
\lefthead{SHIGEYAMA \& TSUJIMOTO}
\righthead{Accretion of dust grains}

\def\etal{{et al. }}
\def\Msun{~M_{\odot} }
\def\cm3{{\rm ~cm}^{-3}}

\def\ltsima{$\; \buildrel < \over \sim\;$}
\def\ltsim{\lower.5ex\hbox{\ltsima}}
\def\gtsima{$\; \buildrel > \over\sim \;$}
\def\gtsim{\lower.5ex\hbox{\gtsima}}

\begin{document}
\title{Accretion of dust grains as a possible origin of metal-poor stars with low $\alpha$/Fe ratios}

\author{Toshikazu Shigeyama$^{1}$, Takuji Tsujimoto$^{2}$ }

\altaffiltext{1}{Research Center for the Early Universe, Graduate
School of Science, University of Tokyo, Bunkyo-ku, Tokyo 113-0033,
Japan; shigeyama@resceu.s.u-tokyo.ac.jp}

\altaffiltext{2}{National Astronomical Observatory, Mitaka-shi,
Tokyo 181-8588, Japan; taku.tsujimoto@nao.ac.jp}

\begin{abstract}
The origin of low $\alpha$/Fe ratios in some metal-poor stars, so called low-$\alpha$ stars, is discussed. It is found that most of low-$\alpha$ stars in the Galaxy are on the main-sequence. This strongly suggests that these stars suffered from external pollution. It is also found that the abundance ratios  Zn/Fe of low-$\alpha$ stars both in the Galaxy and in dwarf spheroidal galaxies are lower than the average value of Galactic halo stars whereas damped Ly $\alpha$ absorbers have higher ratios. This implies that some low-$\alpha$ stars accreted matter depleted from gas onto dust grains. To explain the features in these low-$\alpha$ stars, we have proposed that metal-poor stars harboring planetary systems are the origin of these low-$\alpha$ stars. Stars engulfing a small fraction of planetesimals enhance the surface content of Fe to exhibit low $\alpha$/Fe ratios on their surfaces while they are on the main-sequence, because dwarfs have shallow surface convection zones where the engulfed matter is mixed. After the stars leave the main-sequence, the surface convection zones become deeper to reduce the enhancement of Fe. Eventually, when the stars ascend to the tip of the red giant branch, they engulf giant planets to become low-$\alpha$ stars again as observed in dwarf spheroidal galaxies. We predict that low-$\alpha$ stars with low Mn/Fe ratios harbor planetary systems.
\end{abstract}

\section{INTRODUCTION}
According to the standard Galactic chemical evolution scenario, metal-poor stars have a few times higher abundance ratios of $\alpha$-elements to Fe ($\alpha$/Fe) than those of the sun. This has been ascribed to the fact that massive stars with short life times first contributed to the chemical evolution of the Galaxy by mainly supplying $\alpha$-elements, while type Ia supernovae (SNe Ia) had to wait for their low mass companion stars to turn off the main-sequence to exclusively supply Fe-group elements to the interstellar medium (ISM) \citep[e.g.,][]{Hachisu_99}. As a consequence, stars with [Fe/H]$\gtsim-1$  exhibit decreasing $\alpha$/Fe ratios with increasing Fe/H ratios. On the other hand, stars with [Fe/H]$\ltsim-1$ are expected to have [$\alpha$/Fe]$\sim0.3-0.4$ reflecting yields from massive stars.
Observed spectra of nearby stars have shown a correlation of $\alpha$/Fe ratios with Fe/H supporting this scenario \citep[e.g.,][]{Wheeler_89, Edvardsson_93}.

\begin{figure*}[ht]
\begin{center}
\plotone{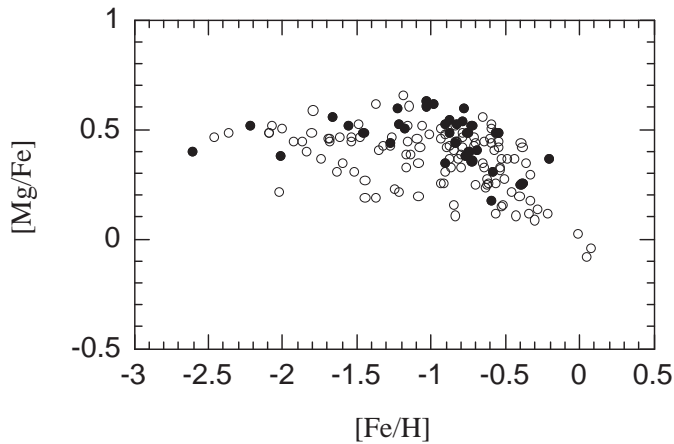}
\caption{The elemental abundance ratios [Mg/Fe] of  field stars \protect{\citep{Gratton_03}} are plotted against [Fe/H].  Open circles represent stars with the surface gravities $\log g>4.1$ and filled circles for stars with $\log g<4.1$.}
\end{center}
\end{figure*}

Recent spectroscopic observations for metal-poor stars have accumulated data exhibiting features of the elemental abundances that are apparently in conflict with the standard scenario: some halo stars exhibit lower [$\alpha$/Fe] values than expected from supernova explosions of massive stars \citep{Nissen_97,Carney_97,King_97,Hanson_98}.  A diversity in [$\alpha$/Fe]  in a large sample of nearby stars with [Fe/H]$\ltsim-1$ (Fig.~1) demonstrates that stars with low [$\alpha$/Fe] ratios are not rare in the Galactic halo \citep{Gratton_03}. A similar diversity in  [$\alpha$/Fe] is also seen in a different large sample of nearby stars observed by \citet{Fulbright_00}.  These data still keep the essential feature that the standard scenario suggests as far as subgiants are concerned (filled circles in Fig.~1). This point will be addressed in the next section.

 The origin of these low-$\alpha$ stars has been discussed in terms of their kinematic parameters \citep{Nissen_97, King_97, Fulbright_00, Gratton_03b}. These authors argued that the $\alpha$/Fe ratios correlated with the apogalactic orbital radius or the perigalactic orbital radius and suggested that the origin of these low-$\alpha$ stars was dwarf galaxies or protogalactic fragments that have undergone a chemical enrichment history different from the rest of the Galactic halo \citep{Nissen_97, Gilmore_98}: such galaxies or fragments might posses an initial mass function (IMF) that enhances the contribution from some massive stars ejecting a large amount of Fe at supernova explosion and/or a burst of star formation that consumes a significant amount of the gas to enhance the later contribution from SNe Ia to the remaining ISM \citep[e.g.,][]{Gilmore_91,Tsujimoto_95a}.

\citet{Shetrone_01} and \citet{Shetrone_03} have revealed detailed elemental abundances of individual stars in nearby dwarf spheroidal (dSph) galaxies and found  that
there exist dSph stars exhibiting low $\alpha$/Fe ratios. Some of them have  ratios even less than the solar value.  In dSph galaxies, a star formation history  different from the Galaxy might lead to earlier contribution from SNe Ia in terms of [Fe/H].  \citet{Ikuta_02} suggested that a very low star formation rate enables SNe Ia to supply Fe to  metal-poor stars with  [Fe/H]$\sim-2$ as observed. However, these stars exhibit [Mn/Fe] ratios as low as $\sim -0.4$ comparable to Galactic halo stars \citep{Shetrone_01b, Shetrone_03}, which indicates that these elements were likely to be synthesized in supernovae originated from massive stars rather than SNe Ia. The Mn/Fe ratio is an important indicator to distinguish products of thermonuclear SNe (SNe Ia) from those of other core-collapse supernovae.  In fact,  theoretical models of SNe Ia have predicted [Mn/Fe]$\sim0$ in the ejecta \citep[e.g.,][]{Nomoto_84, Khokhlov_91}. Hence, if SNe Ia contributed to Mn and Fe in a low-$\alpha$ star, the [Mn/Fe] ratio would become greater than $-0.4$. An IMF different from that of Galactic halo stars would also result in [Mn/Fe] ratios different from $\sim -0.4$.
Furthermore, a short timescale of the star formation inferred from the observed $s$-/$r$-process element ratios strongly suggests that SNe Ia cannot supply heavy elements to dSph  stars. As already discussed in \citet{Tsujimoto_02, Tsujimoto_03}, the observed Ba/Eu (or La/Eu) ratios of dSph stars are very close to the pure $r$-process ratio
\citep{Shetrone_01b,Shetrone_03}, which implies that few of dSph stars had evolved through AGB or  supplied $s$-process elements to other dSph stars till the end of the star formation epoch. Thus the timescale of
the star formation in these dSph galaxies needs to be less than a few
times $10^8$ yr. On the other hand, the progenitor of a SN Ia requires a longer time scale to accrete matter enough to reach the Chandrasekhar mass limit.

In this way, the origin of stars with low $\alpha$/Fe ratios in dSph galaxies is still controversial. In addition, a detailed comparison of elemental abundances of stars in the Galactic halo with those in dSph galaxies casts a doubt on the argument that  low-$\alpha$ stars in the Galactic halo are originated from  low-$\alpha$  stars that have once belonged to dSph galaxies \citep{Fulbright_02}.

In this {\it letter}, we will propose a mechanism to explain the origin of heavy elements in these low-$\alpha$ stars. There are other low-$\alpha$ stars without information on Mn/Fe ratios in young globular clusters Pal 12 and Ruprecht 106 \citep{Brown_97}. The proper motion pair HD 134439/40 also exhibit low $\alpha$/Fe ratios \citep{King_97}, though their Mn/Fe ratios have not been measured. We will also imply the origin of these low-$\alpha$ stars based on our scenario. In the next section,  the surface gravities and effective temperatures of nearby low-$\alpha$ stars are investigated. In \S 3, we discuss the relation between stellar elemental abundances in dSph galaxies and those in damped Ly $\alpha$ (DLA) absorbers. In \S 4, a mechanism is proposed  to explain metal-poor stars with low $\alpha$/Fe ratios both in the Galaxy and in dSph galaxies in a unified manner. In \S 5, conclusions are presented and we discuss some observations to test the proposed mechanism.
\begin{figure*}[ht]
\begin{center}
\plotone{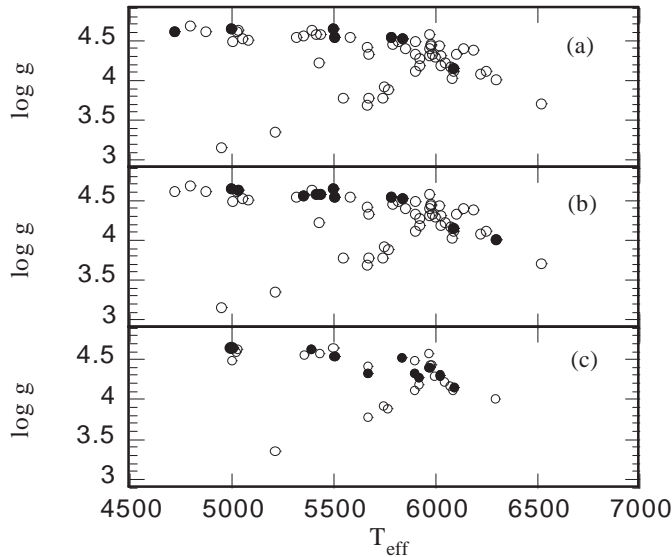}
\caption{The surface gravities ($\log g$) of stars with [Fe/H]$<-1$ taken from  \protect{\citet{Gratton_03}} are plotted against their effective temperatures $T_{\rm eff}$ (Open and filled circles). Filled circles are stars with (a) [Mg/Fe]$<0.3$, (b) [Na/Fe]$<-0.2$, and (c) [Zn/Fe]$<0.05$.}
\end{center}
\end{figure*}

\section{Nearby stars}
If the nearby stars with [Fe/H]$<-1$ observed by \citet{Gratton_03} are plotted in a $\log g-T_{\rm eff}$ plane (Fig.~2a), all the metal-poor stars with [Mg/Fe]$<0.3$ are located in a region limited by $\log g>4.1$ showing that these stars are on the main-sequence. Thus we will refer subgiants to stars with $\log g<4.1$ in the following. On the contrary, stars with [Mg/Fe]$>0.3$ resides both in dwarf and subgiant (red giant) branches. As a result, the elemental abundance pattern of dwarfs in the [Mg/Fe]$-$ [Fe/H] plane (open circles in Fig.~1) have a larger dispersion than that of subgiants. The same feature is seen for other elements like Na and Zn (Fig.~2b, c).  Stars with low Na/Fe and Zn/Fe ratios also reside only on the main-sequence. These results suggest that the abundance features observed in the low-$\alpha$ stars were not solely determined by heavy elements in the ISM from which these stars were formed. Hence some mechanisms in addition to SN nucleosynthesis are needed to explain the observed elemental abundances. In fact, the low-$\alpha$ stars observed by \citet{Gratton_03} have Mn/Fe ratios similar to those of the other halo stars. 

Furthermore, old stars on the main-sequence cannot alter the surface abundances of Mn, Fe, and Zn by nuclear reactions operating inside them.  Therefore the only mechanism to explain the observed feature would be the external pollution that brings some heavy elements into the shallow surface convection zone of a dwarf to significantly alter its surface elemental abundances. For example, the accretion of $\sim0.1\,M_\oplus$ Fe would reduce the abundance ratio [Mg/Fe]  on the surface of a 0.8 $\Msun$ dwarf with [Fe/H]$=-1.5$ by $\sim0.3$ dex.  As the star turns off the main-sequence, the mass of the surface convection zone will increase by a factor of more than 100 and reduce the influence of the external pollution to retrieve the original elemental abundances.

A certain insight into the external pollution introduced here can be obtained by comparing the abundance patterns of Galactic halo stars and DLAs. As shown in Figure 3, the Zn/Fe ratios of DLAs \citep{Prochaska_02} are, on average, larger than those of Galactic halo stars.  Among the halo stars, dwarfs have, on average, lower Zn/Fe ratios than those of subgiants. Since the element Fe is thought to be heavily depleted from gas onto dust grains, the feature of Zn/Fe ratios presented in Figure 3 suggests that the dwarfs with  lower Zn/Fe ratios accreted the Fe that had been depleted from gas whereas higher Zn/Fe ratios in DLAs are results of this Fe depletion. The accretion of the depleted Fe will also explain the feature of Figures 2 since the elements Mg, Zn, and Na are known to be inclined to remain in the gas phase rather than depleted onto dust grains as compared to Fe. 

When Galactic halo stars pass through the Galactic disk, they might accrete the ISM  contaminated by ejecta of SNe Ia. This process might reduce $\alpha$/Fe ratios on the surfaces of stars. However, this must increase Mn/Fe ratios at the same time, leading to the abundance patterns inconsistent with observations. Furthermore, the accretion of the ISM hardly affect $\alpha$/Fe ratios on the surfaces of stars with [Fe/H]$>-2$ because the accretion rate is too small \citep{Yoshii_81}. 

\begin{figure*}[ht]
\begin{center}
\plotone{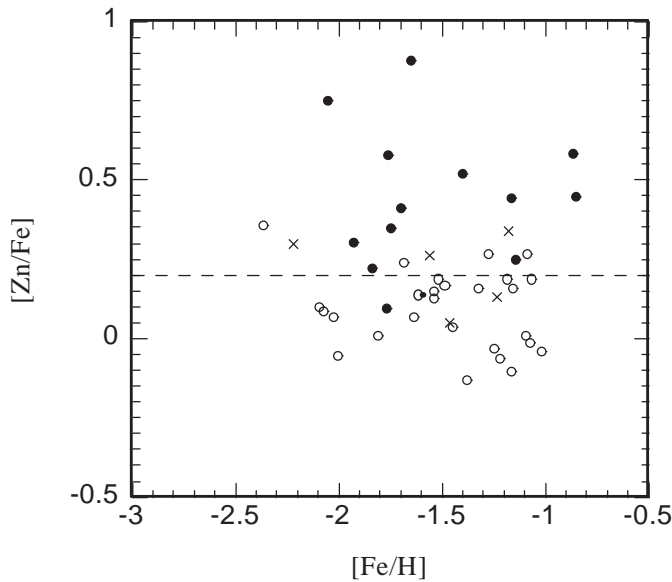}
\caption{The elemental abundance patterns of dwarfs (open circles), subgiants (cross) observed  by \protect{\citet{Gratton_03}}, and DLAs \protect{\citep[open circles;][]{Pettini_00, Prochaska_02}} in the [Fe/H]$-$[Zn/Fe].}
\end{center}
\end{figure*}

\section{Stars in dwarf spheroidal galaxies}
The observed elemental abundances of dSph stars \citep{Shetrone_01b,Shetrone_03} also show  similar relations with those of DLAs for Cr/Fe, Mn/Fe, and Zn/Fe (Fig.~4).  DSph stars have smaller Cr/Fe, Mn/Fe, and Zn/Fe ratios than the average values of Galactic halo stars as well as those of DLAs.  The elements Cr and Mn are also known to be lightly depleted from the gas phase. A sign of the depletion of Fe onto dust grains becomes prominent for  [Fe/H]\gtsim$-2$ in DLAs \citep{Prochaska_02}. On the other hand,  the elemental abundances of dSph stars behave in the same metallicity range as if they accrete Fe that was once depleted onto dust grains. 

As discussed in the preceding section, the surfaces of some nearby low-$\alpha$ stars are affected by dust grains. On the contrary, all the low-$\alpha$ dSph stars showing features of dust grains are very luminous stars residing near the tip of the RGB in the H-R diagram. In the next section, we will propose a mechanism to explain the origin of these low-$\alpha$ stars in different evolutionary stages.

\section{Engulfment of planets}
The envelope of  a star expands as it evolves
along the red giant branch (RGB), and finally the size reaches up to a few AU. Therefore if a star
harbors planets, some of them will be eventually engulfed by their
host star during the late stage of the RGB. For instance, in our solar
system, the sun is expected to eventually engulf all the terrestrial planets, though
the gas giant planets will be beyond the reach of the evolved sun. The ongoing hunting of
extra-solar planets has revealed that there are some planetary systems in which massive planets orbit their host stars with the semi-major axes of the order of only 1 AU. The average mass and semi-major axis of the planet
orbits are found to be $3M_J$ and 1.2 AU in a sample of 107 planets in 93 planetary systems \citep{Schneider_03}, though these estimates probably suffer from the observational bias.  This leads to an implication that some fractions of stars with planets will be likely to engulf giant planets till the stars evolve to the tips of the RGBs.

Though the mass of the dense core in a gas giant planet such as Jupiter and
Saturn in our planetary system is uncertain, a large amount of Fe is
expected to be contained in their cores, compared with terrestrial
planets like Earth \citep{Guillot_99}. It is expected that Jupiter contains roughly five times more Fe than Earth with $0.38M_\oplus$ of Fe \citep[see Table 1
in][]{Murray_01}. Thus a giant planet with the mass of $3M_J$
is expected to contain $\sim 5M_\oplus$ of Fe. This iron mass is in fact
comparable to the mass of Fe in the convection zone of a red giant star
with the mass of $0.8M_\odot$ and the metallicity [Fe/H]$\sim -1.5$.
As a result,  the engulfment of such a giant planet by a metal-poor
red giant star would increase the surface metallicity  by
$\sim +0.3$ dex in [Fe/H] and decrease the Zn/Fe ratio on the stellar surface by a similar amount.

There is no information on how much fraction of each element was depleted from the gas phase in the proto-planetary gas disks around metal-poor stars with the metallicity range of $-2$\ltsim [Fe/H]\ltsim$-1$.  A recent observation of the ISM in the small magellanic cloud
(SMC) with the average metallicity of [Zn/H]$\sim -1$ might, however, shed light on this issue  \citep{Welty_01}: Fe-group elements are depleted as much as in the Galactic ISM, whereas $\alpha$-elements such as Mg and Si are nearly undepleted, though Si has a higher condensation temperature than Fe.  This depletion pattern would support the scenario in which nearby low-$\alpha$ stars engulfed a fraction of dusts in the proto-planetary disks and low-$\alpha$ dSph stars engulfed giant planets.  Of course, the physical conditions could be quite different between proto-planetary gas disks and the observed ISM in the SMC. It is crucial to understand what determines the dust depletion pattern.

The most metal-poor star harboring planets to date is reported to have a metallicity of [Fe/H]=$-0.74$ \citep{Sadakane_02}. Thus the scenario proposed here has not been fully supported by the ongoing search for extra-solar planets. Future observations will construct a much larger sample to answer if there are planets orbiting stars as metal-poor as [Fe/H]$\sim -2$.

\begin{figure*}[ht]
\begin{center}
\plotone{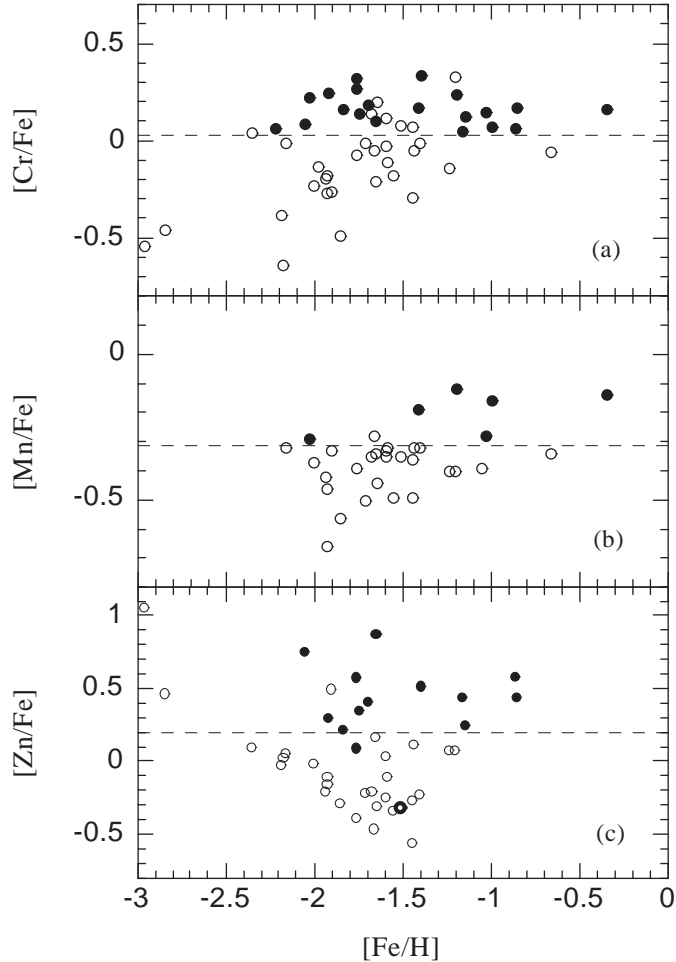}
\caption{The elemental abundance patterns of dSph stars \protect{\citep[open circles;][]{Shetrone_01b, Shetrone_03}} and DLAs \protect{\citep[filled circles;][]{Pettini_00, Prochaska_02}} in the (a)[Cr/Fe]$-$[Fe/H], (b)[Mn/Fe],$-$[Fe/H] and (c) [Zn/Fe]$-$[Fe/H] planes. Dashed lines show the average abundance ratios of Galactic halo stars.}
\end{center}
\end{figure*}

\section{Conclusions and discussion}
We have discussed the origin of low-$\alpha$ stars found in the Galaxy and dSph galaxies. We have shown that the low-$\alpha$ stars found in the Galaxy are on the main-sequence and thus concludes that these stars have suffered from external pollution. A comparison of the elemental abundance pattern of low-$\alpha$ stars  with that of DLAs suggests that low-$\alpha$ stars have accreted dust grains. On the other hand, low-$\alpha$ dSph stars are located at the tip of the RGB.  To explain the origin of these low-$\alpha$ stars in different evolutionary stages, we have proposed that these low-$\alpha$ stars harbor or have harbored planets and that stars engulfing planets and/or planetesimals show low $\alpha$/Fe ratios. However, it should be noted that there is at least one subgiant with low $\alpha$/Fe ratios, BD +80$^\circ$ 245 \citep{Carney_97}. This star has $\alpha$/Fe approximately equal to the solar values and [Fe/H]$\sim -2$. If our scenario is applied to this star, it must have harbored a giant planet at the radius $\ltsim 0.014$ AU. This cannot necessarily discard our scenario.  The smallest semi-major axis of planets in the present catalogue of \citet{Schneider_03} is $\sim 0.02$ AU comparable to the above value. 

Recently,  \citet{Ivans_03} reported  some other metal-poor stars belonging to low-$\alpha$ stars that cannot be explained by the scenario presented here. These stars exhibit extremely large enhancements of Ti, Cr, Mn, Ni, Zn, and Ga relative to Fe and are deficient in Si and Mg.  These elemental abundance features might witness the earliest SNe Ia as suggested by \citet{Ivans_03}.

We cannot expect a correlation of low-$\alpha$ stars that have no sign of SNe Ia with any stellar kinematics from our scenario. Therefore the relationship between low-$\alpha$ stars and stellar kinematics mentioned in \S1 should be explained by some other mechanisms. The correlation of $\alpha/$Fe ratios with the galactic radii was already found in the disk stars \citep{Edvardsson_93}. According to the standard chemical evolution model, this correlation has been explained as a result of the combination of SNe Ia products and longer evolution time scales of the chemical evolution at larger galactic distances \citep{Tsujimoto_95b}. This argument could be applied to the correlation of $\alpha/$Fe ratios with the galactic radii found for the halo stars \citep{Nissen_97, King_97, Fulbright_00, Gratton_03b}. 

It is possible to test our conjecture by two kinds of observations. One is to search planets orbiting around low-$\alpha$ dwarfs in the metallicity range of $-2<$[Fe/H]$<-1$ in the Galaxy. If planets are found to orbit around low-$\alpha$ dwarfs in this metallicity range, it will strongly support our conjecture presented in this {\it letter}.
The other is to search low-$\alpha$ stars with $-2<$[Fe/H]$<-1$ at the tip of the RGB  in the Galaxy. A few spectroscopic observations have been made for red giants with the surface gravities and metallicities similar to the stars observed in dSph galaxies \citep[e.g.,][]{Hanson_98, Fulbright_00}.   None of these stars is a low-$\alpha$ star. The other red giants in the Galaxy observed so far are too metal-poor ([Fe/H]$\ltsim-2$) and/or have too high surface gravities ($\log g\gtsim 1$).

The metallicity distribution function of stars with planets is shown to shift toward the metal rich region compared with that of stars without planets \citep[e.g.,][]{Gonzalez_98}. There have been two explanations for this fact. One is that only stars with sufficient metals can host planets, because the planet formation needs dust grains \citep[e.g.,][]{Santos_03}. The other is that stars with planets tend to engulf planetesimals to enhance their metallicities\citep[e.g.,][]{Gonzalez_98, Murray_02}. Our scenario favors the latter explanation.

\end{document}